\documentclass[a4paper,fleqn]{cas-dc}

\usepackage[numbers]{natbib}
\usepackage{physics, comment, bm,tikz,mwe}
\usetikzlibrary{positioning}

\def\tsc#1{\csdef{#1}{\textsc{\lowercase{#1}}\xspace}}
\tsc{WGM}
\tsc{QE}

\begin{document}
\let\WriteBookmarks\relax
\def\floatpagepagefraction{1}
\def\textpagefraction{.001}

\shorttitle{Flow-Induced Dielectric Axes Rotation in Dipolar Suspensions}    

\shortauthors{P.Srinivasula}  

\title [mode = title]{Weak-Flow Induced Dielectric Axes Rotation in Dipolar Suspensions} 



%

\author[1]{Pramodt Srinivasula}
[orcid=0009-0002-6002-2394]

\cormark[1]


\ead{pramodt.research@gmail.com}



\affiliation[1]{organization={ElectroSoft Labs LLP},
            city={Mumbai},
            country={India}}

\cortext[1]{Corresponding author}



\begin{abstract}
Conventional rheodielectric studies of dipolar suspensions primarily examine flow-induced variations in the principal permittivity components. 
In contrast, an asymptotic solution of the perturbed Fokker--Planck equation for orientable Brownian dipoles under weak flow predicts the emergence of off-diagonal permittivity components that are linear in the relative flow strength. For planar shear flow, these terms exceed the corresponding higher-order diagonal corrections, leading to a rotation of the principal dielectric axes. This previously unrecognized rheodielectric response motivates future experimental validation and suggests new possibilities for flow-controlled dielectric and rheo-optic functionalities.
\end{abstract}

\begin{graphicalabstract}
\includegraphics[width=1\linewidth]{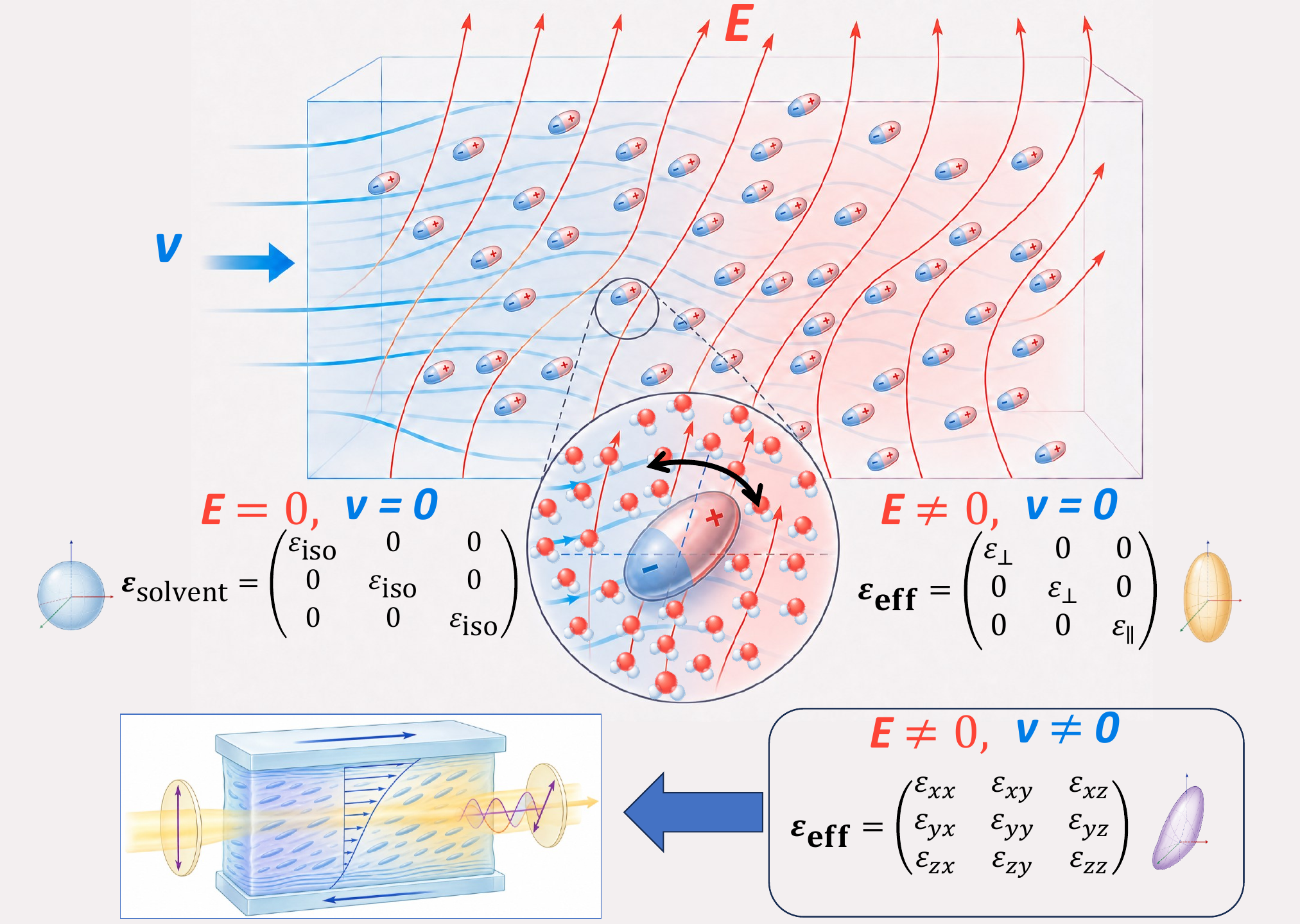}
\end{graphicalabstract}



\maketitle

\section{Introduction}\label{sec:Introduction}

The coupled action of hydrodynamic flow and electric fields governs the orientational dynamics of dipolar entities and thereby the dielectric response of many structured liquids, including polymer solutions, electrorheological fluids, ferrofluids, and colloidal suspensions. While electric-field alignment establishes orientational order, ambient shear perturbs this state, altering the macroscopic polarization and dielectric anisotropy through flow-induced dipole reorientation.

Experimental studies have consistently reported shear-dependent variations in dielectric permittivity across polymer solutions, electrorheological suspensions, and ferrofluids, establishing rheodielectric coupling as a general feature of flowing dipolar media \citep{block1983flow,saimoto1999correlation,adolf1995permittivity,misra2023dichotomous,paulovicova2025rheodielectric}. However, conventional rheodielectric measurements primarily quantify the principal permittivity components, leaving the flow-induced tensorial response, including off-diagonal permittivity and dielectric-axis rotation, largely unexplored.

Theoretical descriptions of dipolar suspensions commonly employ orientational Fokker--Planck equations to relate Brownian orientation dynamics to macroscopic dielectric and rheological properties \citep{raikher1997solution,felderhof1993orientational,coffey2012langevin}. However, analytical solutions remain largely restricted to equilibrium or highly idealized conditions \citep{raikher1997solution}, whereas flow-induced orientational distributions generally require numerical treatments or approximate closures \citep{chinesta2003solution,hu2002significance,cheng2014class,bees1998analytical}. Consequently, closed-form constitutive relations describing the leading tensorial rheodielectric response under coupled electric and flow fields remain unavailable.

To address this gap, in this work, an asymptotic solution of the orientational Fokker--Planck equation for dipolar particles under coupled electric and flow fields, yielding analytical predictions for the leading tensorial rheodielectric response and dielectric-axis rotation.

\section{Methodology}\label{sec:Method}

\begin{figure*}[]
\centering
\begin{tikzpicture}
    \sbox0{\includegraphics[width=0.29\linewidth]{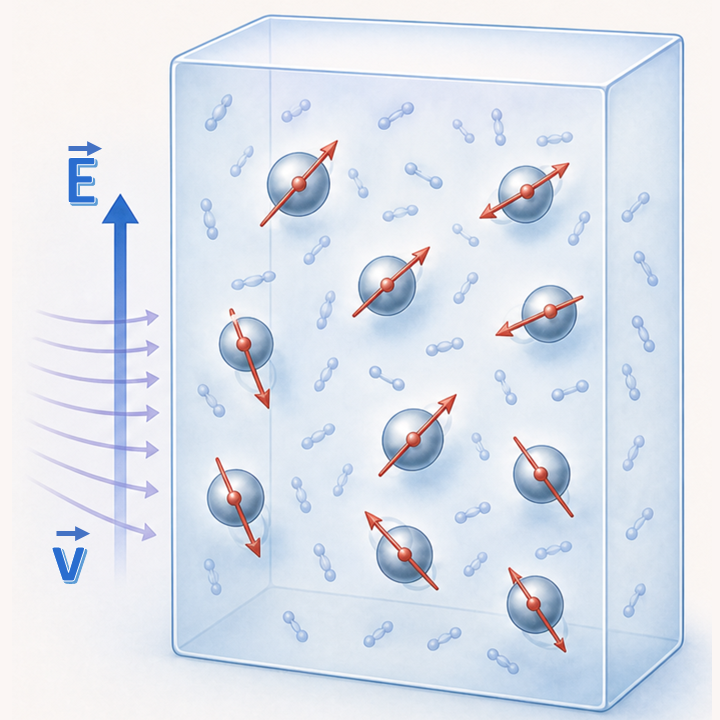} }
    \node[above right,inner sep=0pt] at (0,0)  {\usebox{0}};
    \node[black] at (0.0\wd0,1.0\ht0) {(a) };
\end{tikzpicture}
\begin{tikzpicture}
    \sbox0{\includegraphics[width=0.32\linewidth]{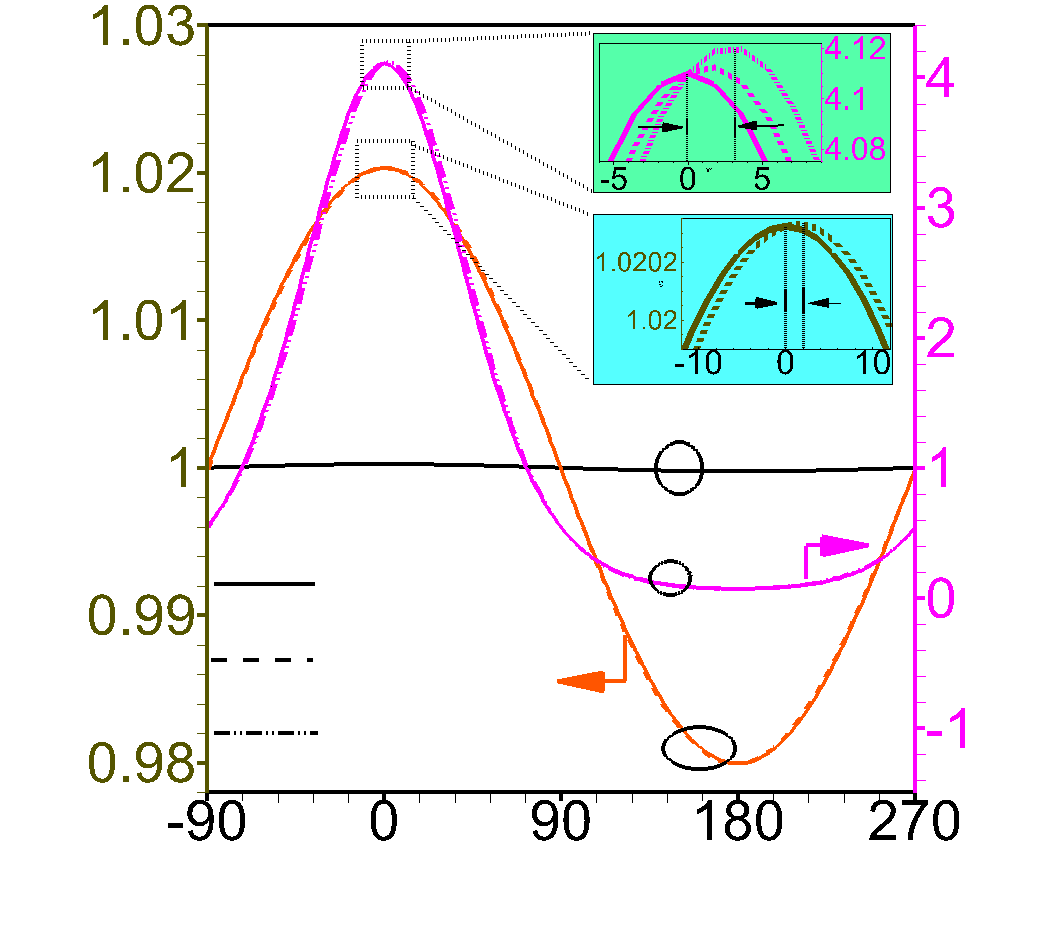}}
    \node[above right,inner sep=0pt] at (0,0)  {\usebox{0}};
    \node[black] at (0.0\wd0,1.0\ht0) {(b) };
    \node[black] at (0.55\wd0,0.04\ht0) {$\vartheta\!=\!\hat{\mathbf{p}}_B \cdot \hat{\mathbf E}$};
    \node[red,rotate=90] at (0.05\wd0,0.5\ht0) {$f\!_B$};
    \node[blue,rotate=270] at (0.95\wd0,0.5\ht0) {$f\!_B$};

    \node[black] at (0.7\wd0,0.54\ht0) {\tiny{$\Lambda_B\!=\!0$}};
    \node[blue] at (0.7\wd0,0.42\ht0) {\tiny{$\Lambda_B\!=\!2$}};
    \node[red] at (0.75\wd0,0.24\ht0) {\tiny{$\Lambda_B\!=\!0.02$}};
    \node[black] at (0.68\wd0,0.85\ht0) {\tiny{$3.6^\circ$}};
    \node[black] at (0.72\wd0,0.65\ht0) {\tiny{$2.2^\circ$}};

    \node[black] at (0.37\wd0,0.39\ht0) {$\alpha\!=\!0$};
    \node[black] at (0.39\wd0,0.31\ht0) {$\alpha\!=\!0.1$};
    \node[black] at (0.39\wd0,0.23\ht0) {$\alpha\!=\!0.2$};
\end{tikzpicture}
\begin{tikzpicture}
    \sbox0{\includegraphics[width=0.32\linewidth]{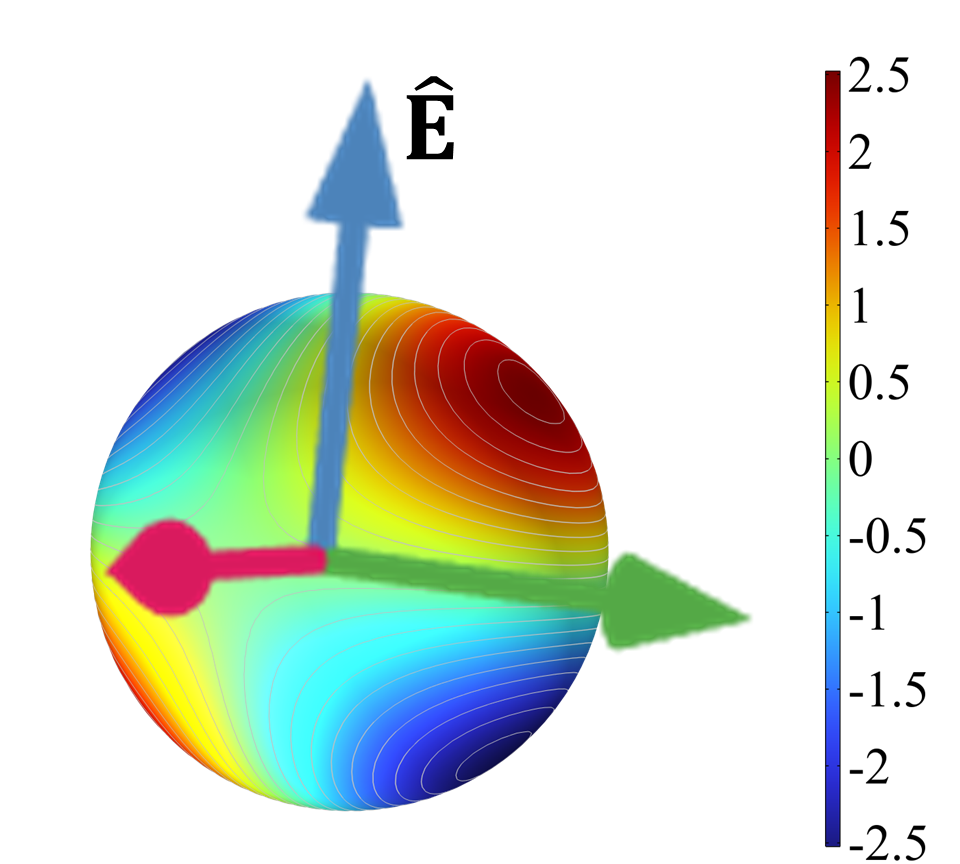}}
    \node[above right,inner sep=0pt] at (0,0)  {\usebox{0}};
    \node[black] at (0.0\wd0,1.0\ht0) {(c) };
    \node[black] at (0.7\wd0,0.97\ht0) {$\left(\frac{f_B-f_B^{(0)}}{f_B^{(0)}}\right)$};    
    \node[black] at (0.9\wd0,0.98\ht0) {$[\%]$};    
\end{tikzpicture}
\caption{\textit{Flow-induced distortion of the orientation distribution of dipolar particles.} (a) Schematic of dipolar particles suspended in a solvent and subjected to concurrent electric and flow fields. (b) Orientation distribution function, $f_B$ (from Eq. \ref{eq:fB_shear_asym}), versus polar angle $\vartheta$ at $\varphi=\pi/2$, illustrating the effects of electric-field strength $\Lambda_B$ and relative flow strength $\alpha$. Inset panels highlight the differences among the curves for the various cases. (c) Three-dimensional representation of the relative flow-induced perturbation on the unit sphere for $\Lambda_B\!=\!2$ and $\alpha\!=\!0.1$, highlighting the symmetry breaking in orientation.}
\label{fig:1_dipole_orientation}
\end{figure*}

\subsection{Thermodynamic description of dipolar suspensions}


To formulate a mathematical description applicable to a broad class of rheodielectric systems, the fluid medium is modeled as a mixture comprising mobile solvent molecules occupying a lattice of fixed total site concentration and suspended orientable Brownian dipolar entities as the dispersed phase. In strongly polar solvents such as water, both the solvent molecules and the suspended dipoles contribute to the polarization response. In contrast, for weakly polar or insulating low dielectric carrier liquids, such as silicone oils, transformer oils, and hydrocarbon-based media, the solvent dipole moment is negligibly small, and the dielectric response is dominated by the suspended dipolar entities. Such a dipolar colloidal suspension subjected simultaneously to electric and flow fields, illustrated schematically in Fig.~\ref{fig:1_dipole_orientation}a, is considered.

The thermodynamic description is formulated through a mean-field free-energy functional $\mathcal{F}$ comprising contributions from the electric field, solvent dipoles, and suspended Brownian dipolar particles. Let $\psi(\mathbf{x},t)$ denote the electric potential. Solvent molecules are modeled as spherical dipoles of effective dipole moment $\mathbf{p}=p_0\hat{\mathbf{p}}$ of fixed magnitude $p_0$ including possible dipolar cavity and intermolecular weak bond effects, with normalized orientational probability density function $f(\hat{\mathbf p};\mathbf{x},t)$. Similarly, the suspended Brownian particles are represented as mildly anisotropic dipolar inclusions with dipole moment $\mathbf{p}_B=p_{0B}\hat{\mathbf{p}}_B$, of magnitude $p_{0B}$, with normalized orientational distribution $f_B(\hat{\mathbf p}_B;\mathbf{x},t)$. The orientational averages are defined as $\langle A\rangle
\!\!=\!\!\int \!\!f A  \, d\hat{\mathbf p}/\int \!\! d\hat{\mathbf p}$ and $\langle A\rangle_B\!\!=\!\!\int\!\! f_B A  \,d\hat{\mathbf p}_B/\int \!\! d\hat{\mathbf p}_B$,
where the integrations are performed over the orientation space of the corresponding dipoles. The total free-energy functional is written as
\begin{eqnarray}
\!\!\mathcal{F}\!\!\!\!\! &=& \!\!\!
\!\!\!\int \!\! -\frac{\varepsilon_0}{2}|\nabla\psi|^2 d\mathbf{x} \! + \mathcal{F}_{s}+ \mathcal{F}_B, \\
\!\!\mathcal{F}_{s}\!\!\!\!\!&=& \!\!\! \!\!\!\int \!\!\!d\mathbf{x}\,c_s\!\left[\!\!- g p_0^2 \!+\!
\left<\!\mathbf{p}\!\cdot\!\nabla \psi \!\right>\!+\! 
 \frac{<\!\ln f\!\!>\!- \lambda_1\!\left(\!<\!\!1\!\!>\!-\!1\!\right)}{\beta}\!\right],\nonumber\\
\!\mathcal{F}_{B}\!\!\!\!\!&=&\!\!\!\!\!\!
\!\int \!\!\!d\!\mathbf{x} c_B\!\!\left[\!
\left<\mathbf{p}_{B}\!\cdot\!\nabla \psi \right>_B \!+  \!\frac{<\!\ln f_B\!>_B\!- \lambda_{1B}\!\left(<\!1\!>_B\!-\!1\!\right)}{\beta}\!\right]\!\!. \nonumber
\label{eq:totalFreeEnergy}
\end{eqnarray}
Here, $\varepsilon_0$ is the permittivity of free space and $\beta\!=\!(k_BT)^{-1}$ is the inverse thermal energy. The parameter $g$ denotes the reaction-field associated with the solvent dipoles in a polar solvent, while $c_s$ and $c_B$ are the number concentrations of solvent molecules and concentration of Brownian particles. The first term in $\mathcal F$ represents the electrostatic self-energy of the electric field. The solvent contribution $\mathcal F_s$ accounts for reaction-field interactions, dipole--field coupling, and orientational entropy of the solvent dipoles. The Brownian contribution $\mathcal F_B$ contains the corresponding dipole--field interaction and orientational entropy of the suspended particles. The Lagrange multipliers $\lambda_1(\mathbf{x})$ and $\lambda_{1B}(\mathbf{x})$ enforce normalization of the orientational distribution functions, $\int \!\!f\,d\hat{\mathbf p}\!=\!4\pi$ and $\int \!\! f_B \,d\hat{\mathbf p}_B\!=\!4\pi$, respectively. This decomposition explicitly separates the orientational energetic contributions of the solvent and suspended dipolar species.


The suspended dipolar particles are modeled as weakly anisotropic prolate spheroids whose dipole moments are aligned with their major axes. The particle shape is characterized by the aspect ratio $\Theta_B\!>\!1$, with the corresponding Jeffery shape factor $\chi_B\!\!=\!\!((\Theta_B^2\!-\!1)\!/\!(\Theta_B^2\!+\!1) )$  \citep{perrin1934mouvement,koenig1975brownian}.
Let $\boldsymbol{\kappa}\!\!=\!\!\nabla\mathbf{v}$ denote the velocity-gradient tensor of the ambient flow, decomposed into its antisymmetric and symmetric parts as $\boldsymbol{\kappa}\!=\!\mathbf{W}\!+\!\mathbf{D}$, with anti-symmetric vorticity tensor $\mathbf{W}\!\!=\!\!(\boldsymbol{\kappa}-\boldsymbol{\kappa}^T)/2$ and symmetric rate of deformation tensor $\mathbf{D}\!\!=\!\! (\boldsymbol{\kappa}+\boldsymbol{\kappa}^T)/2$.
Following Jeffery's theory, the flow-induced reorientation rate of the particle orientation vector is
\begin{equation}
\dot{\hat{\mathbf p}}_{B,f}=\hat{\hat{\mathbf p}}_B^{\perp}
\cdot
\left(
\mathbf W
+
\chi_B \mathbf D
\right)
\cdot
\hat{\mathbf p}_B,
\label{eq:JefferyOrientation}
\end{equation}
where $\hat{\hat{\mathbf p}}_B^{\perp}\!=\!\mathbf I\!-\!\hat{\mathbf p}_B\hat{\mathbf p}_B$ is the projection operator onto the plane normal to $\hat{\mathbf p}_B$ \citep{kim2013microhydrodynamics}. Equation~(\ref{eq:JefferyOrientation}) describes the combined effects of local fluid rotation and strain-induced alignment on the suspended dipoles, while preserving the unit-magnitude constraint $|\hat{\mathbf p}_B|\!=\!1$. The hydrodynamic reorientation rate $\dot{\hat{\mathbf p}}_{B,f}$ subsequently enters the orientational dynamics of the distribution function $f_B(\hat{\mathbf p}_B;\mathbf{x},t)$ together with the electric-field-induced torque and rotational diffusion.

The orientational dynamics of the suspended dipolar particles are derived using Onsager's variational principle. The total dissipation functional is written as the sum of viscous dissipation in the carrier fluid and rotational dissipation associated with the Brownian particles. The solvent contribution is
\begin{equation}
\Phi_s=\int d\mathbf{x}\, \eta_f
\left( \boldsymbol{D} \mathbin{:} \boldsymbol{D} \right), 
\end{equation}
where $\eta_f$ is the viscosity of the ambient fluid and $\boldsymbol{\kappa}\!=\!\nabla\mathbf{v}$ is the velocity-gradient tensor \citep{doi2012onsager}.
Following Doi--Onsager theory for orientable anisotropic particles \citep{doi1988theory,beris2024dissipation}, the rotational dissipation is expressed as
\begin{equation}
\Phi_B=\!\!\int \!\! d\mathbf{x}\,
\frac{c_B\zeta_B^r}{2} \!
\left\langle \!
(\dot{\hat{\mathbf p}}_B-\dot{\hat{\mathbf p}}_{B,f})^2
\! + \!
\frac{\chi_B}{2}
(\hat{\mathbf p}_B\cdot\boldsymbol{D}\cdot\hat{\mathbf p}_B)^2
\!\right \rangle\!{_{_B}} ,
\end{equation}
where $\zeta_B^r$ is the rotational friction coefficient and $\dot{\hat{\mathbf p}}_{B,f}$ is the flow-induced reorientation rate given by Eq.~(\ref{eq:JefferyOrientation}). The first term represents dissipation due to the relative rotation between the particle and the surrounding flow, whereas the second term represents the additional viscous dissipation associated with extensional strain along the major axis of the rigid spheroidal particles. For weak anisotropy ($\delta_B\!=\!\Theta_B\!-\!1\!\ll\!1$), the rotational friction coefficient $\zeta_B^r$ is taken to be isotropic to leading order \citep{satohintroduction}.

The Onsager variational functional (Rayleighian) is then defined as
\begin{equation}
\mathcal R=\Phi_s
+
\Phi_B
+
\dot{\mathcal F}-\int\!\!\!\int d\mathbf{x} d\hat{\mathbf p}_B\,
\lambda_2\left(\dot{\hat{\mathbf p}}_B\cdot\hat{\mathbf p}_B\right),
\label{eq:Rayleighian}
\end{equation}
where the rate of change of the free-energy functional is
\begin{equation}
\dot{\mathcal F}=\int d\mathbf{x}
\left(\frac{\delta\mathcal F}{\delta f}\dot f + \frac{\delta\mathcal F}{\delta f_B}\dot f_B
+ \frac{\delta\mathcal F}{\delta\psi}\dot\psi\right),
\end{equation}
the last term with Lagrange multiplier $\lambda_2$ enforces the unit-magnitude constraint $|\hat{\mathbf p}_B|\!=\!1$ and $\delta \mathcal F \!/\! \delta ( \cdot )$ denotes the variational derivative.

\subsection{Fokker--Planck model for particle distribution}

The equilibrium orientational distribution of the solvent dipoles is obtained by minimizing the free-energy functional with respect to $f$, yielding the classical Langevin--Boltzmann distribution. The equilibrium orientational distribution $f\!=\!(\Lambda \tilde{E}/sinh(\Lambda  \tilde{E})) e^{\Lambda \hat{\mathbf{p}} \cdot \tilde{\mathbf{E}}}$, is obtained from $\delta \mathcal{F}\!/\!\delta f\!=\!0$ \citep{iglivc2010excluded}. Here $\tilde{\mathbf{E}}\!=\!\tilde{E}\hat{\mathbf{E}}\!=\!-\nabla\psi/E_0$ is a normalized electric field along the unit vector $\hat{\mathbf{E}}$, for a characteristic field magnitude of $E_0$, and $\Lambda\!=\! p_{0}E_0\beta$ is a dimensionless number indicating relative magnitude of electric field influence on the molecular dipolar orientation. 

In contrast, the suspended Brownian dipoles remain dynamically coupled to both the electric field and the ambient flow and therefore govern the flow-dependent dielectric response.
Conservation of orientational distribution on the unit sphere requires
\begin{equation}
\frac{\partial f_B}{\partial t}=-\nabla_{\hat{\mathbf p}_B}
\cdot
\left(
f_B\dot{\hat{\mathbf p}}_B
\right),
\label{eq:orientation_conservation}
\end{equation}
where $\nabla_{\hat{\mathbf p}_B}$ denotes the gradient operator in orientation space. The orientational dynamics of the Brownian particles are obtained from Onsager's variational principle of minimization of the Rayleighian, $\delta\mathcal R/\delta\dot{\hat{\mathbf p}}_B\!=\!0$,
\begin{equation}
\dot{\hat{\mathbf p}}_B=\dot{\hat{\mathbf p}}_{B,f}-
D_B^r \hat{\hat{\mathbf p}}_B^{\perp} \cdot
\left(
\beta p_{0B}\nabla\psi
+
\nabla_{\hat{\mathbf p}_B}
\ln f_B
\right),
\label{eq:orientation_velocity}
\end{equation}
where $D_B^r\!=\!(\beta\zeta_B^r)^{-1}$ is the rotational diffusivity of the particles.
Substituting Eq.~(\ref{eq:orientation_velocity}) into Eq.~(\ref{eq:orientation_conservation}) gives the governing Fokker--Planck equation
\begin{equation}
\frac{\partial f_B}{\partial t}\!=\!\nabla_{\hat{\mathbf p}_B} \!\cdot \left[\!  D_B^r  \hat{\hat{\mathbf p}}_B^{\perp} \!\cdot\!
\left( \beta p_{0B} f_B\nabla\psi + \nabla_{\hat{\mathbf p}_B}f_B\right)\!
-\! f_B\dot{\hat{\mathbf p}}_{B,f}
\!\right].
\label{eq:FP}
\end{equation}
Equation~(\ref{eq:FP}) describes the competition between electric-field-induced alignment, rotational diffusion, and hydrodynamic reorientation by the ambient flow. This orientational Fokker--Planck equation constitutes the central kinetic model used in the subsequent asymptotic analysis of polarization and flow-modified dielectric permittivity.

\section{Results and Analysis}
\subsection{Weakly-advected dielectric corrections}

Motivated by rheodielectric conditions where electric-field alignment and rotational diffusion relax the orientational distribution much faster than hydrodynamic advection, the steady-state orientational Fokker--Planck equation (Eq. \eqref{eq:FP}) may be integrated once in orientation space, yielding a constant vector contribution. As orientational transport is confined to the tangent space of the unit sphere, this contribution identically vanishes. In terms of the microscopic definition of the Mason number is $\alpha\!=\! \overline{\alpha}/\Lambda_B \!=\! \dot{\gamma} \chi_B/(D_B^r\Lambda_B)$, with a characteristic strain rate $\dot{\gamma}$, the dimensionless electric field influence on dipolar particle orientation $\Lambda_B\!=\! p_{0B} E_0 \beta$ and the particle Péclet number $\overline{\alpha}$ we obtain,
\begin{equation}
\nabla_{\hat{\mathbf p}_B}f_B
=
f_B \Lambda_B
\hat{\hat{\mathbf p}}^{\perp}_B
\cdot
\left(
\tilde{\mathbf E}
+
\alpha \tilde{\boldsymbol D}\cdot\hat{\mathbf p}_B
\right).
\label{eq:reduced_fp}
\end{equation}
Here  $\tilde{\mathbf{D}} \!=\! \mathbf{D}/\dot{\gamma}$ is the normalized deformation tensor. Integrating Eq. (\ref{eq:reduced_fp}) and enforcing the normalization $\!\int\!\!f_B\, d\hat{\mathbf{p}}_B\!=\!4\pi$ yields
\begin{equation}
f_B
=
\frac{4\pi}{Z}
\exp
\left[
\Lambda_B
\left(
\hat{\mathbf p}_B\cdot\tilde{\mathbf E}
+
\frac{\alpha}{2}
\hat{\mathbf p}_B
\cdot
\tilde{\mathbf D}
\cdot
\hat{\mathbf p}_B
\right)
\right],
\end{equation}
with a partition function,
\begin{equation}
Z
=
\int
\exp
\left[
\Lambda_B
\left(
\hat{\mathbf p}_B\cdot\tilde{\mathbf E}
+
\frac{\alpha}{2}
\hat{\mathbf p}_B
\cdot
\tilde{\mathbf D}
\cdot
\hat{\mathbf p}_B
\right)
\right]
d\hat{\mathbf p}_B.
\end{equation}

For a weak flow coupling, $\alpha \! \ll \!\! 1$ of an incompressible fluid ($tr(\tilde{\mathbf{D}})\!=\!0$) with a finite $\Lambda_B \!\gtrsim\! 1$, $f_B$ can be approximated asymptotically in terms of Langevin function $\mathcal L (A)\!=\!coth(A)-1/A$ as, 
\begin{align}
\!\!\!\!\!\!\!\!\!\!\! \!\!\!\!\!
f_B  = & f_{B}^{(0)} + \alpha f_{B}^{(1)} + O(\alpha^2), \label{eq:fB_asymp}\\
\!\!\!\!\!\!\!\!\!\!\! \!\!\!\!\!
f_B^{(1)}\!\!=& \frac{\Lambda_B f_{B}^{(0)}}{2}  \!\!\Big[
\hat{\mathbf p}_B  \!\!\cdot\!  \tilde{\mathbf D} \!\cdot\!  \hat{\mathbf p}_B \!-  \!\!\left(\!\!  1  \!\!+\!  
\frac{\mathcal L(\beta_B)}{\beta_B} \!
-\! \frac{2}{(\beta_B)^2} \!\! \right) \!\!
\left( \! \hat{\mathbf E}  \!\cdot\!  \tilde{\mathbf D}  \!\cdot\!  \hat{\mathbf E}  \!\right)\!\! 
\Big]\!
\end{align}
with $\beta_B\!=\!\Lambda_B \tilde E$ and the equilibrium orientation distribution $f_{B}^{(0)} \!=\!\frac{\beta_B e^{\beta_B(\hat{\mathbf p}_B\cdot\hat{\mathbf E})} }{\sinh(\beta_B)}$.
The advective perturbation modifies the field-aligned orientational microstructure through a strain-induced quadrupolar redistribution of particle orientations, with the distortion magnitude governed by the local rate-of-strain tensor while maintaining, $\int \!\! f_B^{(1)} d\hat{\mathbf p}_B\!=\! 0$.


Further, using Eq. \ref{eq:fB_asymp}, the total polarization $\mathbf{P}\! = \! \mathbf{P}_{s} \!+ \mathbf{P}_{B}$, with solvent and suspended-particle contributions $\mathbf{P}_{s} \!=\! c_s \!\left< \mathbf{p} \right> $ and $\mathbf{P}_{B} \!=\! c_B\! \left< \mathbf{p}_B \right>\!_B $, respectively, is evaluated as,
\begin{align*}
\mathbf{P}_{s} &=  c_s p_0 \mathcal{L}(\Lambda\tilde E) \,\hat{\mathbf E}, \\
\mathbf{P}_{B} &=  \mathbf{P}_{B}^{(0)} + \alpha \mathbf{P}_{B}^{(1)} + \mathcal{O}(\alpha^2) ,\\
\mathbf{P}_{B}^{(1)} &= \mathbf{P}_{B}^{(0)} \left[  A \tilde{\mathbf D}\cdot\hat{\mathbf E} + B (\hat{\mathbf E}\cdot\tilde{\mathbf D}\cdot\hat{\mathbf E}) \hat{\mathbf E} \right],\\
A
&=
\frac{1}
{\tilde E\,\mathcal L(\Lambda_B\tilde E)}
-
\frac{3}
{\Lambda_B\tilde E^2}, \\
B &= \frac{15}{2\Lambda_B\tilde E^2} - \frac{5}  {2\tilde E\,\mathcal L(\Lambda_B\tilde E)}
+ \frac{3}{2} \frac{\mathcal L(\Lambda_B\tilde E)}{\tilde E}.
\end{align*}
Here $\mathbf{P}_{B}^{(0)} \!=\!  c_B p_{0B} \mathcal{L}(\Lambda_B \tilde E) \,\hat{\mathbf E} $ is the contribution of the particulate to the equilibrium polarization without advection.
If $\tilde{\mathbf D}\!\cdot\!\hat{\mathbf E}
=(\hat{\mathbf E}\!\cdot\!\tilde{\mathbf D}\!\cdot\!\hat{\mathbf E})\,\hat{\mathbf E}$, the polarization remains collinear with the electric field. Thus, when the electric field is aligned with a principal strain direction, the flow modifies only the polarization magnitude. Otherwise, the polarization acquires a transverse component and is no longer aligned with the electric field.
Hence, the total polarization vector density is, 
\begin{align}
\!\!\!\!\!\!\!\!\!\!\!\!
\mathbf P &= \mathbf P^{(0)} + \alpha \mathbf P^{(1)} + O(\alpha^2), \label{eq:P_general}\\ 
\!\!\!\!\!\!\!\!\!\!\!\!
\mathbf P^{(0)} \!&=\! M\!(\tilde E)\hat{\mathbf E}, \, 
\mathbf P^{(1)} \!=\! \Big[\! R(\tilde E) \,\tilde{\mathbf D} \!\cdot\! \hat{\mathbf E} + S(\tilde E)  (\hat{\mathbf E}\cdot\tilde{\mathbf D} \!\cdot\! \hat{\mathbf E}) \hat{\mathbf E} \!\Big], \nonumber \\
\!\!\!\!\!\!\!\!\!\!\!\!
M(\tilde E) &= c_s p_0 \mathcal L(\Lambda\tilde E) + c_B p_{0B} \mathcal L(\Lambda_B\tilde E),\nonumber\\ 
\!\!\!\!\!\!\!\!\!\!\!\!
R(\tilde E) &= c_B p_{0B} \mathcal L(\Lambda_B\tilde E) A, \quad S(\tilde E) = c_B p_{0B} \mathcal L(\Lambda_B\tilde E)B .\nonumber
\end{align}

The relative permittivity tensor is defined as $(\varepsilon_r)_{ij}=\delta_{ij} + \frac{1}{\varepsilon_0} \frac{\partial P_i}{\partial E_j}$ is hence estimated as,
\begin{equation}
\boldsymbol{\varepsilon}_r = \boldsymbol{\varepsilon}_r^{(0)} + \alpha \boldsymbol{\varepsilon}_r^{(1)} + \mathcal O(\alpha^2).
\label{eq:eps_expansion}
\end{equation}
with
\begin{equation}
\boldsymbol{\varepsilon}_r^{(0)}  = \mathbf I + \frac{M}  {\varepsilon_0 E_0 \tilde E} \left(\mathbf I - \hat{\mathbf E}\hat{\mathbf E} \right) 
+ \frac{M'} {\varepsilon_0 E_0} \hat{\mathbf E}\hat{\mathbf E}.
\label{eq:eps0_tensor}
\end{equation}
and 
\begin{align}
\!\!\!\!\!\! \!\! \!\! \!
\boldsymbol{\varepsilon}_r^{(1)} \! = \! 
\frac{1}{\varepsilon_0 E_0} \!\!
& \Bigg[ \!
R'
(\tilde{\mathbf D}\!\cdot\!\hat{\mathbf E})
\hat{\mathbf E}  \!+\!
\frac{R}{\tilde E}\,
\tilde{\mathbf D}
\!\cdot\!
\left(
\mathbf I
\!-\!
\hat{\mathbf E}\hat{\mathbf E}
\right) \!+\!
S'(\hat{\mathbf E}\!\cdot\! \tilde{\mathbf D} \!\cdot\!\hat{\mathbf E})
\hat{\mathbf E}\hat{\mathbf E}
\nonumber \\
\!\!\!\! \!\! \!\! \!\! \!
& + \frac{S}{\tilde E} 
\Big[
2 \hat{\mathbf E} (\tilde{\mathbf D} \!\cdot\!\hat{\mathbf E})
+ (\hat{\mathbf E}\!\cdot\! \tilde{\mathbf D} \!\cdot\!\hat{\mathbf E}) 
\left( \mathbf I - \hat{\mathbf E}\hat{\mathbf E} \right) \Big] \Bigg],
\label{eq:eps1_tensor}
\end{align}
where the notations $M',R'$ and $S'$ imply their derivatives with $\tilde E$, respectively.

Therefore, Eqs.~\eqref{eq:eps0_tensor} and \eqref{eq:eps1_tensor} provide the quiescent dielectric response and the first-order flow-induced correction to the relative permittivity tensor, respectively. Note the terms with the tensor components $\hat{\mathbf E}\hat{\mathbf E}$ and $\Big( \mathbf I \!-\! \hat{\mathbf E}\hat{\mathbf E} \Big)$ in Eq.~\eqref{eq:eps0_tensor} represent the tensor components parallel ($\varepsilon_{r \parallel}$) and orthogonal ($\varepsilon_{r \perp}$) to the electric field. 
The order $\mathcal O (\alpha)$ flow-induced dielectric correction in Eq.~\eqref{eq:eps1_tensor} scales linearly with the strain-rate tensor and is governed by two geometric invariants: the longitudinal projection $\hat{\mathbf E} \!\cdot\! \tilde{\mathbf D} \!\cdot\! \hat{\mathbf E}$, which modifies the principal dielectric response, and the transverse projection $(\mathbf I \!-\! \hat{\mathbf E}\hat{\mathbf E})\!
\cdot\! \tilde{\mathbf D} \!\cdot\! \hat{\mathbf E}$, which generates off-diagonal dielectric coupling. Consequently, the perturbation with off-diagonal components arising whenever the electric-field direction is not an eigenvector of the strain-rate tensor.

\subsection{Planar flow in rheodielectric experiments}

\begin{figure*}[]
    \centering
\begin{tikzpicture}
    \sbox0{\includegraphics[width=0.45\linewidth]{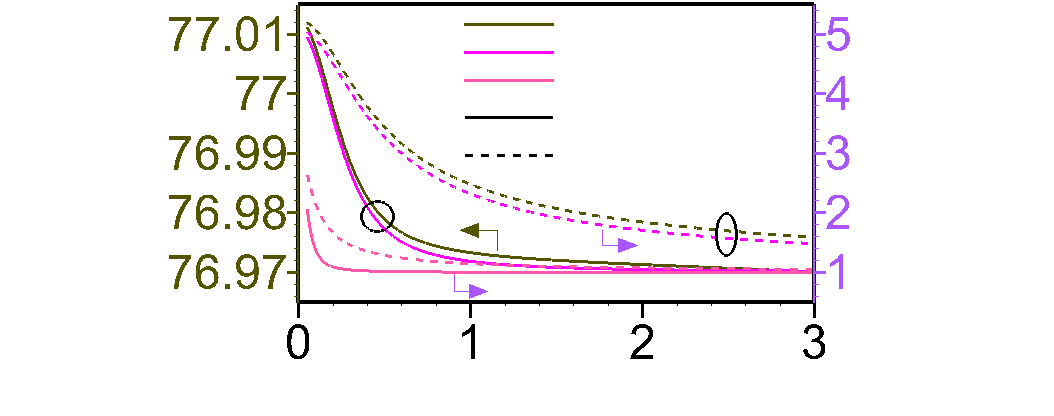}}
    \node[above right,inner sep=0pt] at (0,0)  {\usebox{0}};
    \node[black] at (0.0\wd0,1.0\ht0) {(a) };
    \node[black] at (0.53\wd0,0.03\ht0) {$E_0$ (MV/m)};
    \node[red,rotate=90] at (0.1\wd0,0.64\ht0) {$\varepsilon_{\parallel},\varepsilon_{\perp}$};
    \node[blue,rotate=270] at (0.85\wd0,0.64\ht0) {$\varepsilon_{\parallel},\varepsilon_{\perp}$};
    \node[black] at (0.58\wd0,0.95\ht0) {\small{polar}};
    \node[black] at (0.63\wd0,0.88\ht0) {\small{nonpolar-1}};
    \node[black] at (0.63\wd0,0.8\ht0) {\small{nonpolar-2}};
    \node[black] at (0.56\wd0,0.71\ht0) {\small{$\varepsilon_\parallel$}};
    \node[black] at (0.56\wd0,0.61\ht0) {\small{$\varepsilon_\perp$}};
    \node[black] at (0.4\wd0,0.48\ht0) {$\varepsilon_\parallel$};
    \node[black] at (0.71\wd0,0.51\ht0) {$\varepsilon_\perp$};
\end{tikzpicture}
\begin{tikzpicture}
    \sbox0{\includegraphics[width=0.42\linewidth]{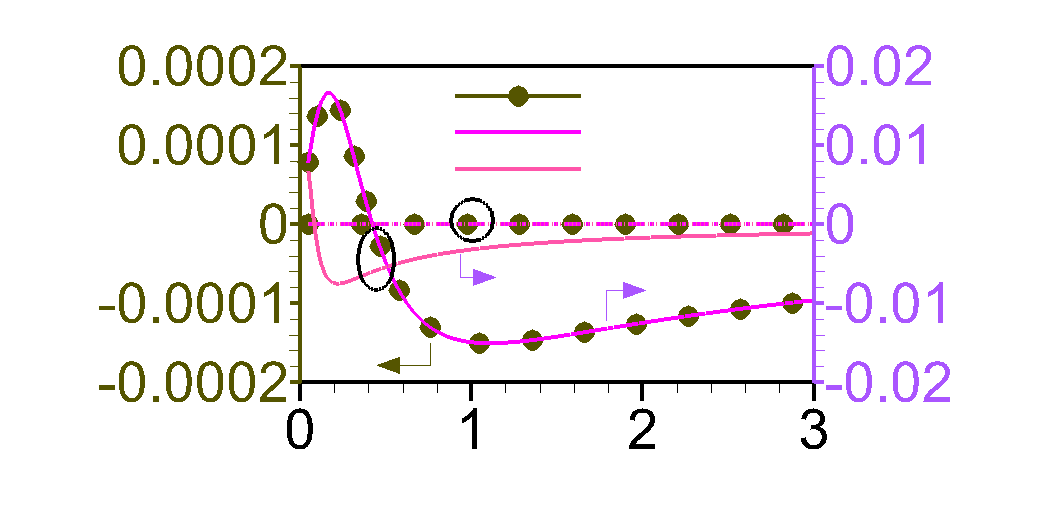}}
    \node[above right,inner sep=0pt] at (0,0)  {\usebox{0}};
    \node[black] at (0.0\wd0,0.87\ht0) {(b) };
    \node[red,rotate=90] at (0.05\wd0,0.64\ht0) {$\varepsilon_{\theta \theta}$};
    \node[blue,rotate=270] at (0.95\wd0,0.64\ht0) {$\varepsilon_{\theta \theta}$};
    \node[black] at (0.61\wd0,0.82\ht0) {\small{polar}};
    \node[black] at (0.66\wd0,0.75\ht0) {\small{nonpolar-1}};
    \node[black] at (0.66\wd0,0.67\ht0) {\small{nonpolar-2}};
    \node[black] at (0.345\wd0,0.38\ht0) {$\alpha\!=\!0.1$};
    \node[black] at (0.4\wd0,0.64\ht0) {$\alpha\!=\!0$};
    \node[black] at (0.53\wd0,0.03\ht0) {$E_0$ (MV/m)};
\end{tikzpicture}
    \caption{\textit{Flow-induced perturbation of the permittivity tensor and dielectric-axis rotation.} (a) Principal components of the relative permittivity tensor in the absence of flow $\alpha\!=\!0$ and (b) flow-induced off-diagonal component, $\varepsilon_{\theta z}$, as functions of $E_0$ (Eq. \ref{eq:epsr_shear}) for representative polar and nonpolar suspensions listed in Table~1.}
    \label{fig:2_permittivity}
\end{figure*}

To demonstrate the applicability of the present theoretical framework to experimentally relevant systems, three representative cases are selected to span polar and nonpolar solvents and different particle concentrations. In the literature, rheodielectric measurements have been performed using parallel-annulus \citep{adolf1995permittivity}, coaxial-cylinder \citep{saimoto1999correlation}, and parallel-plate \citep{misra2023dichotomous} geometries, with the imposed shear flow and electric field mutually perpendicular. \citet{adolf1995permittivity} studied $1~\mu$m ceramic particles in nonpolar dodecane at $10$--$30\%$ volume fractions, $0.2$--$1$ kV/mm at 400 Hz, and $0.5$--$200$ s$^{-1}$ steady shear rate. \citet{saimoto1999correlation} investigated silica particles in nonpolar silicone oil at $0.1$--$0.3$ volume fractions, $1$--$4$ kV/mm, with measurements spanning 30--1000 Hz and a higher $150$--$1400$ s$^{-1}$ shear rate. \citet{misra2023dichotomous} studied PNIPAM hydrogel particles in water using $\sim1.5$ V/mm in a wide range of frequency of 2Hz--5MHz and oscillatory shear rates of approximately $1$--$200$ s$^{-1}$. Order-of-magnitude estimates of the corresponding model parameters give $\Lambda_B\sim3500$, $1$--$10$, and $10^{-4}$ for the Adolf, Saimoto, and Misra systems, respectively. These experimentally motivated ranges therefore provide a basis for selecting the three representative cases considered here, as summarized in Table~1.

To illustrate the general theory, a planar shear flow, typical to the experiments with annulus ring rheometer, is considered in cylindrical coordinates \((r,\theta,z)\) of basis $(\mathbf e_r, \mathbf e_\theta, \mathbf e_z)$, with velocity field \(v_r=v_z=0\) and \(v_\theta=(\Omega r/h)z\), yielding a local shear rate \(\dot{\gamma}(r)=\Omega r/h\). The electric field is axial, \(\hat{\mathbf E}=\mathbf e_z\) and the planar shear flow considered here, $\tilde{\mathbf D} = \left( \mathbf e_\theta\mathbf e_z + \mathbf e_z\mathbf e_\theta \right)\!/4 $. 
The orientation vector in spherical coordinates, polar angle $\vartheta$ and azimuthal angle $\varphi$ of a particle, $\hat{\mathbf p}_B \!=\! \sin\vartheta\cos\varphi\mathbf e_r\! +\! \sin\vartheta\sin\varphi\mathbf e_\vartheta \!+\! \cos\vartheta\mathbf e_z$ the asymptotic normalization constant becomes, $Z = 4\pi \sinh(\Lambda_B\tilde E)/\Lambda_B\tilde E + O(\alpha^2)$. Hence the orientational distribution can be simplified as,
\begin{align}
\!\!\!\!\!\!\!\!\!\!\!\!\!\!
f_B(\vartheta,\varphi)
=
&\frac{\Lambda_B\tilde E}
{\sinh(\Lambda_B\tilde E)}
e^{\Lambda_B\tilde E\cos\vartheta}
\left[
1
+
\frac{\alpha\Lambda_B}{8}
\sin(2\vartheta)\sin\varphi
\right]\nonumber \\
&+
O(\alpha^2).
\label{eq:fB_shear_asym}
\end{align}
The variation of $f_B$ with the polar angle $\vartheta$ at $\varphi=\pi/2$ for different values of $\Lambda_B$ and $\alpha$ is shown in Fig.~\ref{fig:1_dipole_orientation}b. As $\Lambda_B$ increases, the orientational distribution becomes progressively concentrated about the electric-field direction. Under weak shear, peak of the distribution undergoes a small angular displacement of approximately $2^\circ$--$4^\circ$ away from the field direction, for $\Lambda_B \!\sim\! 0.02$--$2$. The corresponding perturbation, shown in Fig.~\ref{fig:1_dipole_orientation}c, reveals a maximum correction of approximately $\pm 2.5\%$, occurring away from the electric-field axis, thus breaking the symmetry.

On the other hand, a measurement of polarization or permittivity does not resolve the orientation of each individual Brownian dipole. Instead, it observes the collective response of ensemble of dipoles with the microscopic orientational degree of freedom homogenized, leaving a macroscopic dielectric response controlled by the electric and flow fields. The polarization density Eq. \ref{eq:P_general} reduces to,
\begin{align}
\mathbf P = M(\tilde E)\,\mathbf e_z + \alpha C(\tilde E)\,\mathbf e_\theta + O(\alpha^2),\\
C(\tilde E) = \frac{c_B p_{0B}}{4} \left[
\frac{1}{\tilde E} - \frac{3\mathcal L(\Lambda_B\tilde E)} {\Lambda_B\tilde E^2} \right].
\end{align}
The constitutive relation for the relative permittivity is specialized to an imposed axial electric field, so the permittivity is evaluated only through $\partial\mathbf{P}/\partial E_z$. Therefore, the asymptotic relative permittivity tensor becomes
\begin{equation}
\boldsymbol{\varepsilon}_r
=
\begin{pmatrix}
1+\dfrac{M}{\varepsilon_0E_0\tilde E}
&
0
&
0
\\[2ex]
0
&
1+\dfrac{M}{\varepsilon_0E_0\tilde E}
&
\alpha\dfrac{C'(\tilde E)}
{\varepsilon_0E_0}
\\[2ex]
0
&
0
&
1+\dfrac{M'}{\varepsilon_0E_0}
\end{pmatrix}
+
\mathcal O(\alpha^2),
\label{eq:epsr_shear}
\end{equation}
with, $C'(\tilde E) =\dv{C}{\tilde E}$, indicating that the planar shear flow does not modify the diagonal dielectric response at \(O(\alpha)\). Instead, it induces an off-diagonal \((\theta,z)\) component in the permittivity tensor, reflecting the flow-induced transverse polarization generated by the Brownian dipolar particles. 

Table~\ref{tab:parameters} summarizes experimentally relevant parameters for polar \citep{misra2023dichotomous} and nonpolar \citep{adolf1995permittivity,saimoto1999correlation} dipolar particle suspensions, used to evaluate the permittivity tensor variations shown in Fig.~\ref{fig:2_permittivity}(a,b). The diagonal components, $\varepsilon_{zz}\!=\!\varepsilon_{\parallel}$ and $\varepsilon_{rr}\!=\!\varepsilon_{\theta\theta}\!=\!\varepsilon_{\perp}$, decrease and gradually saturate with electric-field strength as the orientational freedom of the dipoles decreases, while remaining unchanged by the flow at $\mathcal{O}(\alpha)$. In contrast, a non-zero off-diagonal component, $\varepsilon_{\theta z}$, emerges under weak shear ($\alpha\!=\!0.1$), exhibiting distinct electric-field dependence for the polar and nonpolar suspensions.

\begin{table}[t]
\centering
\caption{Representative material parameters employed for the illustrative calculations of polar and nonpolar dipolar suspensions.}
\label{tab:parameters}
\begin{tabular}{|p{2.25cm}|p{1.4cm}|p{1.45cm}|p{1.45cm}|}
\hline
Parameter & Polar & Nonpolar$\,1$ & Nonpolar$\,2$ \\
\hline
Solvent number density,
$c_s$ ($\mathrm{m^{-3}}$)
&
$3.35\times10^{28}$
&
$6.0\times10^{27}$
&
$6.0\times10^{27}$
\\

\hline
Solvent dipole moment,
$p_0$ (D)
&
4.5
&
$1.0\times10^{-2}$
&
$1.0\times10^{-2}$
\\

\hline
Particle number density,
$c_B$ ($\mathrm{m^{-3}}$)
&
$4.0\times10^{18}$
&
$4.0\times10^{20}$
&
$8.0\times10^{18}$
\\

\hline
Particle dipole moment,
$p_{0B}\,$(D)
&
$1.0\times10^{4}$
&
$1.0\times10^{4}$
&
$5.0\times10^{4}$
\\

\hline
\end{tabular}
\end{table}

An eigenvalue analysis of the asymptotic permittivity tensor in Eq. \ref{eq:epsr_shear} quantifies the reorientation of the principal dielectric axes induced by shear flow. The off-diagonal \(O(\alpha)\) correction preserves the principal permittivities but rotates the eigenvector associated with the axial dielectric response, resulting in a flow-dependent tilt ($\Theta$) of the dielectric principal direction in the ($\theta,z$)-plane. 
\begin{equation}
\tan\Theta = \frac{\alpha C'(\tilde E)}{M'\tilde E-M}.
\label{eq:eigenvector_tilt}
\end{equation}
The emergence of flow-induced off-diagonal components renders the permittivity tensor non-diagonal in the laboratory coordinate system. Consequently, its orthogonal principal dielectric axes rotate relative to those of the zeroth-order tensor.
The asymptotic expression up to order $\alpha$ for the dielectric-axis rotation is valid away from parameter regimes where $M'(\tilde E) \tilde E \!-\! M(\tilde E)$ approaches zero, since the corresponding eigenvalue splitting becomes vanishingly small and higher-order corrections may become non-negligible, i.e., for small $\Lambda_B$. On the other limiting value as $\Lambda_B$ increases, $\Theta$ reaches a limiting value of $(\alpha/4\tilde E^2)$, for small tilt $\Theta\!\ll\!1$ and strong field effect $\Lambda_B\!\gg\!1$.
\begin{figure*}[]
    \centering
\begin{tikzpicture}
    \sbox0{\includegraphics[width=0.42\linewidth]{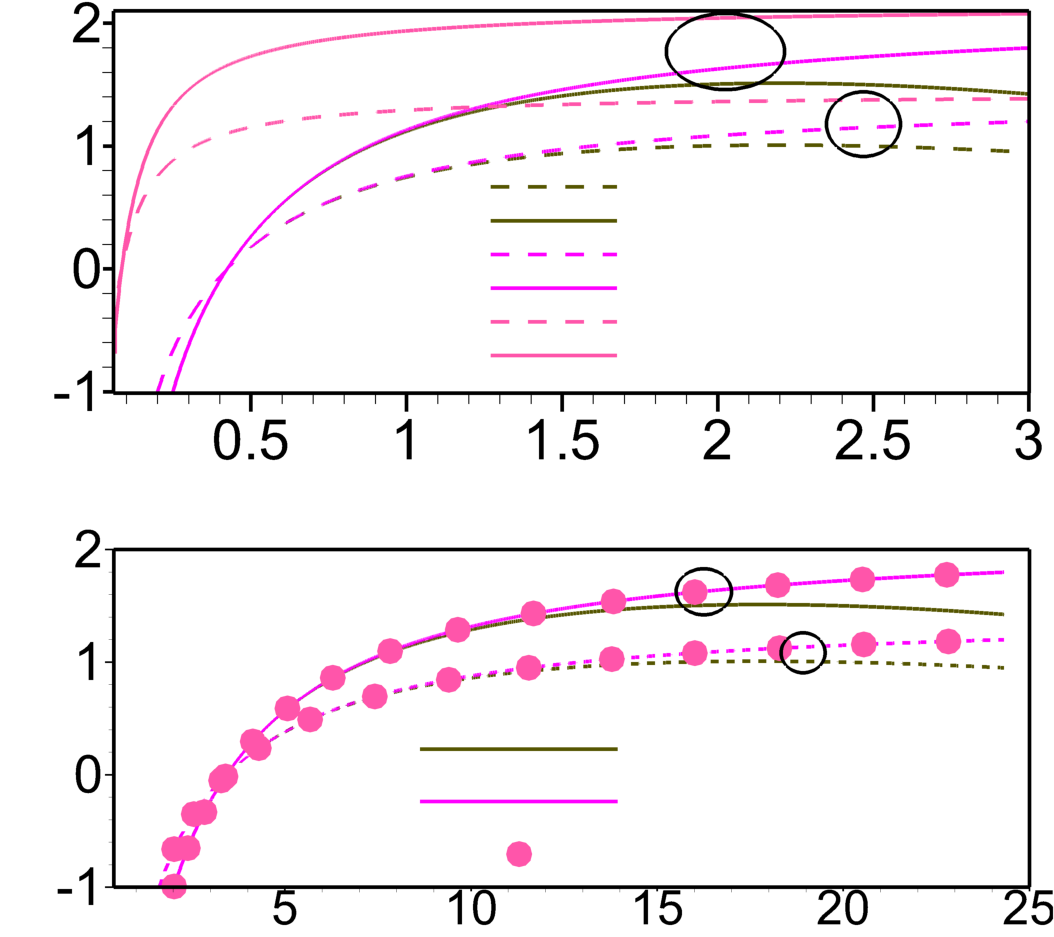} }
    \node[above right,inner sep=0pt] at (0,0)  {\usebox{0}};
    \node[black] at (0.0\wd0,1.0\ht0) {(a) };
    \node[black] at (0.58\wd0,0.47\ht0) {$E_0$ (MV/m)};
    \node[black,rotate=90] at (0.02\wd0,0.8\ht0) {$\Theta$};
    \node[black] at (0.0\wd0,0.425\ht0) {(b) };
    \node[black,rotate=90] at (0.02\wd0,0.25\ht0) {$\Theta$};
    \node[black] at (0.57\wd0,-0.02\ht0) {$\Lambda_B$};
    
    \node[red] at (0.71\wd0,0.81\ht0) {\small{polar $\alpha\!=\!0.1$}};
    \node[red] at (0.72\wd0,0.77\ht0) {\small{polar $\alpha\!=\!0.15$}};
    \node[blue] at (0.76\wd0,0.735\ht0) {\small{nonpolar-1 $\alpha\!=\!0.1$}};
    \node[blue] at (0.77\wd0,0.7\ht0) {\small{nonpolar-1 $\alpha\!=\!0.15$}};
    \node[purple] at (0.76\wd0,0.66\ht0) {\small{nonpolar-2 $\alpha\!=\!0.1$}};
    \node[purple] at (0.77\wd0,0.62\ht0) {\small{nonpolar-2 $\alpha\!=\!0.15$}};
    
    \node[red] at (0.65\wd0,0.21\ht0) {\small{polar}};
    \node[blue] at (0.7\wd0,0.15\ht0) {\small{nonpolar-1}};
    \node[purple] at (0.7\wd0,0.09\ht0) {\small{nonpolar-2}};

     \node[black] at (0.55\wd0,0.95\ht0) {\small{$\alpha\!=\!0.15$}};
    \node[black] at (0.92\wd0,0.87\ht0) {\small{$\mathbf{\alpha\!=\!0.1}$}};
    \node[black] at (0.56\wd0,0.39\ht0) {\small{$\alpha\!=\!0.15$}};
    \node[black] at (0.78\wd0,0.27\ht0) {\small{$\alpha\!=\!0.1$}};
\end{tikzpicture}
\begin{tikzpicture}
    \sbox0{\includegraphics[width=0.46\linewidth]{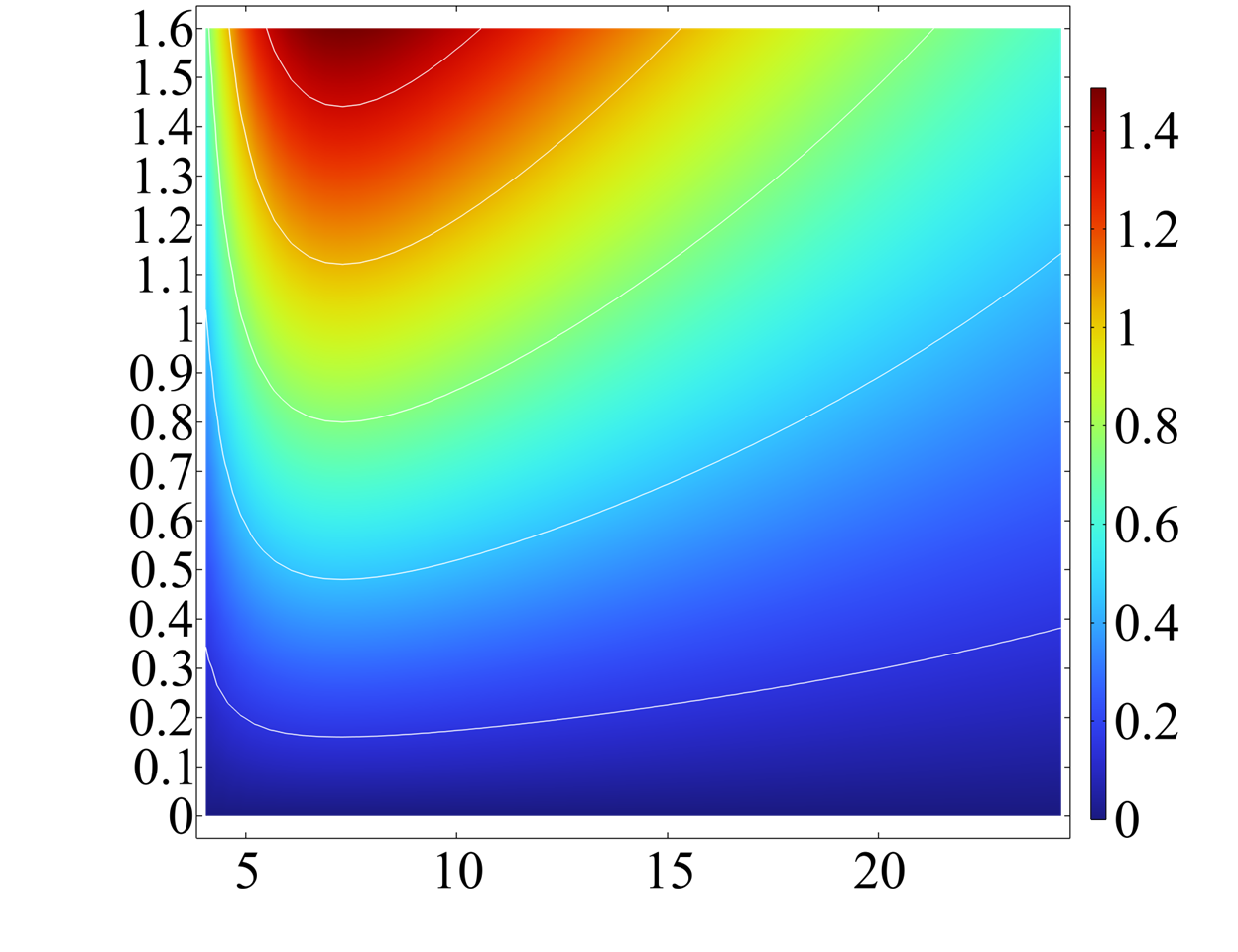}}
    \node[above right,inner sep=0pt] at (0,0)  {\usebox{0}};
    \node[black] at (0.0\wd0,1.0\ht0) {(c) };
    \node[black] at (0.5\wd0,1.04\ht0) {$\Theta$ map};
    \node[black] at (0.94\wd0,0.94\ht0) {$\Theta$ ($^\circ$)};
    \node[black] at (0.57\wd0,0.0\ht0) {$\Lambda_B$};
    \node[black] at (0.05\wd0,0.55\ht0) {$\overline \alpha$};
    
    \node[black] at (0.3\wd0,0.86\ht0) {$1.34^{\circ}$};
    \node[black] at (0.52\wd0,0.85\ht0) {$1.04^{\circ}$};
    \node[black] at (0.3\wd0,0.6\ht0) {$0.74^{\circ}$};
    \node[black] at (0.55\wd0,0.44\ht0) {$0.45^{\circ}$};
    \node[black] at (0.75\wd0,0.26\ht0) {$0.15^{\circ}$};
\end{tikzpicture}
    \caption{\textit{Flow-induced perturbation of the dielectric-axis rotation.} (a) Dielectric-axis rotation angle, ($\Theta$), as a function of the dimensional electric field strength ($E_0$), and (b) the corresponding nondimensional field strength ($\Lambda_B$), from Eq. \ref{eq:eigenvector_tilt} for representative cases listed in Table~1. (c) Contour map of $\Theta$ in the $\overline{\alpha}$--$\Lambda_B$ parameter space, with colors and contour lines indicating the coupled dependence on the electric and flow fields.}
    \label{fig:3_dielec_rotation}
\end{figure*}

Figures~\ref{fig:3_dielec_rotation}a and \ref{fig:3_dielec_rotation}b show the variation of the dielectric-axis rotation angle $\Theta$ with the electric-field strength and the corresponding dimensionless parameter $\Lambda_B$, respectively, for $\alpha=0.1$ and $0.15$. The $\Theta$ increases monotonically with both the electric-field strength and the Mason number $\alpha$, reflecting the increasing influence of hydrodynamic advection relative to electric-field alignment. Furthermore, when expressed in terms of the dimensionless parameters $\alpha$ and $\Lambda_B$, the results for both polar and nonpolar suspensions collapse onto nearly identical curves. 
The tilt $\Theta$ obtained from Eq.~\ref{eq:eigenvector_tilt} can be expressed in terms of the dimensionless parameters as
\begin{align}
&\!\!\!\!\!\!\!\!\!\!\!\!\!\!\!\!\tan\Theta = \nonumber \\ 
&\!\!\!\!\!\!\!\!\!\!\!
\frac{\alpha}{4\tilde E^2}
\frac{
-1-3\mathcal L'(\Lambda_B\tilde E)
+\dfrac{6\mathcal L(\Lambda_B\tilde E)}
{\Lambda_B\tilde E}
}{
\sigma \!\!
\left[\!
\Lambda\tilde E\mathcal L'\!(\Lambda\tilde E)
\!-\! \mathcal L(\Lambda\tilde E)\!
\right]
\!\!+\!\!
\Lambda_B\!\tilde E\mathcal L'\!(\Lambda_B \!\tilde E)
\!-\! \mathcal L(\Lambda_B \!\tilde E)
},\!\!
\label{eq:tan_theta_dimless_full}
\end{align}
where $\sigma \!=\! (c_sp_0)/(c_Bp_{0B})$ quantifies the relative contribution of the solvent to the dipolar response of the suspension. This contribution can become more important with increasing solvent polarity or decreasing particle concentration. However, the molecular dielectric response of the solvent is typically weak, such that $\Lambda \!\ll\! 1$ \citep{srinivasula2025dipolar}. In this limit, the solvent contribution to the denominator of Eq.~\ref{eq:tan_theta_dimless_full} becomes negligible of the order $\sigma\Lambda^3 \!\ll\! 1$. Consequently, $\Theta$ depends primarily on the particle-related dimensionless parameters $\alpha$ and $\Lambda_B$ (in addition to the dimensionless field strength $\tilde E \!\sim\! \mathcal O(1)$).

The corresponding contour map in the $(\overline{\alpha},\Lambda_B)$ Péclet--dipolar strength parameter space for small Mason numbers, $\alpha\lesssim0.1$, is shown in Fig.~\ref{fig:3_dielec_rotation}c for the polar suspension. The map illustrates the coupled dependence of the dielectric-axis rotation on the independently controlled electric-field strength $\Lambda_B$ and shear rate $\overline \alpha$, thereby providing a direct connection to experimentally accessible conditions. Notably, the dielectric tilt exhibits a non-monotonic dependence on $\Lambda_B$ due to its influence on both the principal and off-diagonal components of the permittivity tensor. An optimal dipolar strength $\Lambda_B$ maximizes the flow-induced rotation. Moreover, this optimal $\Lambda_B$ shifts toward higher values with increasing shear rate $\overline{\alpha}$.

\section{Discussion}

Shear flow exerts a hydrodynamic torque that competes with electric-field alignment, producing a steady tilt of the mean dipole orientation away from the field direction. This flow-induced orientational distortion propagates from microscopic dipole dynamics to the polarization density and ultimately to the macroscopic dielectric response. For weak shear, the perturbation changes the orientational distribution antisymmetrically about the electric-field direction, so that the transverse polarization survives whereas the longitudinal correction cancels upon orientational averaging. Consequently, the effective permittivity develops off-diagonal components while its principal permittivities remain unchanged up to terms proportional to $\alpha$. The dielectric tensor therefore rotates its principal axes without an accompanying $\mathcal O(\alpha)$ change in its principal values, indicating that weak shear primarily breaks dielectric symmetry through orientation redistribution rather than changes in polarization magnitude.

Hydrodynamically, the rheodielectric response is governed by the combined action of the particle shape factor $\chi_B$ and the strain-rate tensor $\tilde {\mathbf{D}}$, which together generate the flow-induced orientational distortion. The dispersed phase ranges from nearly spherical ceramic particles \citep{adolf1995permittivity} to highly anisotropic (rod-like) constituents. The present theory is inline with the observations that increase in particle anisotropy toward elongated particles ($\chi_B \!\rightarrow\! 1$) exhibit increasing electro-rheological and rheo-dielectric responses \citep{dong2019electrorheological,jekal2025comparative,kovaleva2024electrorheological}. However, the present theory is developed in the limit of small anisotropy ($\delta_B\! \ll\! 1$) and presented the results for demonstrative cases.
Dielectrically, the formulation is independent of the microscopic mechanism responsible for the effective dipole moment, making it applicable to orientable dipolar entities whose polarization may originate from permanent molecular polarity, or other coarse-grained effective dipolar mechanisms. However, many practical systems may consist of more complex dipolar constituent mechanisms, especially with strong induced or interfacial polarizations, and hence may systematically deviate from the current predictions.

The present asymptotic solution targets the weak-shear regime, which is common in nanofluidic systems, while providing the foundation for systematic higher-order extensions. At $\mathcal O(\alpha^2)$, quadratic flow--field couplings generate higher-order orientational moments, introducing additional tensorial contributions to both the principal and off-diagonal dielectric responses. Such higher-order effects are expected to become increasingly important under the stronger flow conditions. The analysis further considers suspensions where hydrodynamic and electrostatic interactions among the particles can be neglected, and represents the steady-state solution of the Brownian orientational dynamics of the particles.

The present formulation is sufficiently general to be incorporated into arbitrary flow configurations, providing a constitutive framework for continuum modeling while enabling future integration with rheo-optical characterization techniques. While conventional parallel-plate rheodielectric measurements are primarily sensitive to the principal permittivity components ($\varepsilon_\parallel,\varepsilon_\perp$), the predicted off-diagonal response suggests new opportunities for experimental validation; such as, through low-frequency electrical rheodielectric measurement with polarization tensor-sensitive geometric setup which is not widely available today for validation. 

The frequency of the applied AC voltage in rheodielectric experiments is typically varied to probe polarization mechanisms characterized by distinct time scales, spanning low- and high-frequency relaxation processes \citep{misra2023dichotomous} and very-low-frequency responses associated with conductive charge migration in nonpolar solvents \citep{paulovicova2025rheodielectric}. In electrolytic systems, the reorientation of molecular dipoles may additionally couple to the evolution of the electric double layer, introducing another relevant time scale \citep{srinivasula2026dipolar}. In contrast, the present work predicts a flow-modified static (zero probing-frequency) dielectric tensor, obtained from the steady orientation distribution governed by the Fokker--Planck equation. The predictions can therefore be tested under steady shear using a low-frequency AC probe that approaches the quasi-static dielectric response, while avoiding, or independently calibrating, slower parasitic contributions such as electrode polarization and ionic migration not captured by the present orientational model. 

A suitably designed multi-electrode geometry could, in principle, extend rheodielectric measurements to the transverse electrical response required to resolve the off-diagonal components of the flow-modified response. Generalized spectroscopic ellipsometry has demonstrated reconstruction of complete dielectric tensors from polarization-sensitive measurements performed at different sample orientations \citep{sturm2015dielectric}. More recently, dielectric tensor tomography has enabled direct reconstruction of three-dimensional dielectric tensors in nematic materials by combining illumination at different angles and polarization states with measurements of the resulting vectorial optical fields \citep{shin2022tomographic}. These approaches, however, probe the optical dielectric response and therefore do not directly access the low-frequency response considered here. Nevertheless, their underlying strategy of acquiring multiple independent polarization responses to reconstruct the dielectric tensor, beyond a reduced scalar quantity, provides a useful conceptual basis for developing an analogous low-frequency electrical measurement approach. In particular, replacing the optical excitation and field detection with multi-directional low-frequency electrical excitation and voltage/current measurements under steady shear could, in principle, enable reconstruction of the complex admittivity tensor and thereby provide access to the off-diagonal dielectric response after accounting for conductive contributions.

Extending the present formulation to a time-dependent perturbation of the Fokker--Planck equation would provide a natural route to predicting the amplitude and phase of the dielectric response, thereby connecting the evolving dipolar orientation to the full complex dielectric response measured at finite AC frequencies. Such a formulation could directly probe frequencies comparable to the characteristic rotational relaxation rate, estimated here as $f_c \!\sim\! D_B^r/(2\pi)$. In the regime where Brownian rotational relaxation and shear-induced reorientation occur on comparable time scales, oscillatory shear could provide an additional means of probing the dynamic orientational response. This would extend the present steady-state theory toward a frequency-dependent rheodielectric description and might as well enable leveraging optical axis rotation measurement techniques \citep{carey2024giant,zhao2003optical} for determining the dielectric rotation dynamics under shear.

Beyond validation, the present results provide a theoretical basis for exploiting flow-controlled dielectric anisotropy in rheodielectric and rheo-optical characterization of electro-responsive soft materials. Whereas a diagonal dielectric response modifies the strength of an existing electrical response, the emergence of tensorial permittivity introduces a cross-directional electric displacement response to an applied electric field in a weak flow. This flow-induced coupling between orthogonal electrical and mechanical responses may give rise to nontrivial electrohydrodynamic phenomena, particularly in confined micro- and nanofluidic systems.

\section{Conclusions}\label{sec:Conclusions}

While rheodielectric experiments have long focused on shear-induced variations in the principal dielectric permittivity, the accompanying flow-induced tensorial response has remained largely inaccessible to conventional measurement configurations. In this work, a first-principles analytical framework was developed by asymptotically solving the orientational Fokker--Planck equation for dipolar particles under coupled electric and flow fields. The analysis shows that weak planar shear tilts the dipolar orientation away from the electric-field direction, generating off-diagonal dielectric responses proportional to the Mason number ($\alpha$), which quantifies the ratio of hydrodynamic to electric torques; whereas changes in the principal permittivities arise only at higher asymptotic orders. Consequently, dielectric-axis rotation emerges as the leading rheodielectric signature of weak flow-induced anisotropy. 
The resulting constitutive framework quantitatively links orientational dynamics to the dielectric tensor in a general, coordinate-independent formulation, providing a foundation for continuum modeling, tensor-sensitive rheodielectric experiments, the design of flow-responsive dielectric materials, and the discovery of new electrohydrodynamic and optofluidic phenomena.

\textit{\textbf{Declarations}--} No data were created or used in this study. This research received no specific grant from any funding agency. The author declares no competing interests.

\bibliographystyle{cas-model2-names}

\bibliography{cas-refs}



\end{document}